\documentstyle[prl,aps]{revtex}
\input epsfig.sty
\begin{document}
\draft
 
\title{Bose-Einstein Condensates in the Large Gas Parameter Regime} 

\author{
A.\ Fabrocini$^{({\rm a,b})}$, and
A.\ Polls$^{({\rm c})}$}
\address{$^{({\rm a})}$INFN, Sezione di Pisa, I-56100 Pisa, Italy}
\address{$^{({\rm b})}$Department of Physics, University of Pisa, I-56100 Pisa, Italy}
\address{$^{({\rm c})}$Departament d'Estructura i Constituents de la Mat\`eria, 
Universitat de Barcelona, E-08028 Barcelona, Spain}
\date{\today}
\maketitle
\begin{abstract}
Bose--Einstein condensates of 10$^4$ $^{85}$Rb atoms in a cylindrical 
trap are studied using a recently proposed modified Gross-Pitaevskii 
equation. The existence of a Feshbach resonance allows for widely 
tuning the scattering length of the atoms, and values of the 
peak gas parameter, $x_{pk}$, of the order of 10$^{-2}$ can be attained. 
We find large differences between the results of the modified 
Gross-Pitaevskii and of the standard Thomas--Fermi, and Gross-Pitaevskii 
equations in this region. The column densities at $z=0$ may differ by as 
much as $\sim 30\%$ and the half maximum radius  by $\sim 20\%$. 
The scattering lengths estimated by fitting the half maximum radius  
within different approaches  can differ by $\sim 40\%$. 
\end{abstract}

Bose-Einstein condensation (BEC) of magnetically trapped alkali 
atoms has been achieved in several experimental setups, most of them 
in regimes where the atomic gas is considered to be very dilute,
i.e. the average interatomic distance is much larger than the range of 
the interaction. As a consequence, the physics is dominated by two 
body collisions, generally well described in terms of the $s$--wave 
scattering length, $a$. The crucial parameter defining the condition 
of diluteness is the gas parameter, $x({\bf r})=n({\bf r}) a^3$,  
where $n({\bf r})$ is the local density. 
For low values of the average gas parameter, 
$x_{av}\leq 10^{-3}$, the mean field Gross-Pitaevskii (GP) 
equation~\cite{Pita61} is the logical tool to study the system.

The gas parameter can be brought  outside the regime of validity of the GP 
equation by two ways: either by increasing the number of atoms, $N$,  
in the condensate, or by changing their effective size. 
 Recent experiments have explored both possibilities. 
On one side they have reached very large $N$ values, $N\sim 10^8$;  
on the other the scattering lengths have been widely tuned. 
The second approach promises to be much more efficient to reach 
large $x$ regions. 

In a recent experiment performed at JILA~\cite{Cornish1,Today}, it was 
possible to confine about $10^4$ atoms of $^{85}$Rb in a cylindrical 
trap. By exploiting the presence of a Feshbach resonance at 
a magnetic field  of $B\sim 155$ Gauss, the scattering length 
was varied from negative to very high positive values. Actually, it 
was suggested some time ago~\cite{Stoof1} that $a$ 
could be modulated by taking advantage of its expected strong variation 
in the vicinity of a magnetic induced Feshbach resonance in collisions 
between cold alkali atoms. Several experiments have supported 
this proposal by demostrating this type of scattering length variation  
for several alkali atoms, as $^{85}$Rb, Cs and Na~\cite{Roberts,Vuletic,Inouye}. 
However, only very recently the Boulder group at JILA has been succesful 
in producing stable $^{85}$Rb condensates where  
$a$ can be effectively tuned over a very wide range.  

To explore with confidence such regimes of high values of the parameter 
$x$, a necessary task is to investigate  the accuracy of the GP equation. 
Moreover, to quantify the limitations of the GP description is 
of particular relevance since the empirical estimate of the scattering 
length is based upon the so called Thomas-Fermi (TF) approximation 
to the GP equation. The TF approximation amounts to disregard the 
kinetic energy term in the GP equation. The TF and GP results are expected to 
 coincide for large $N$ and/or $a$ values. 
In the experimental analysis, a common procedure 
consists in measuring the column density, given by the integral of the 
particle density along a direction perpendicular to the simmetry 
axis of the trap, $n_c(z) =\int dx~n(x,0,z)$. The $x$  
direction coincides with that of the light beam used to image 
the atomic cloud. Then, the scattering length is inferred by finding 
the value of $a$ that, in the framework of the TF equation, provides a 
 column density with  the same experimental 
 size.

In this paper, we will use a recently proposed modified Gross Pitaevskii
equation (MGP)~\cite{Fabro99} to estimate the corrections to the GP results. 
We begin by briefly recalling the derivation of the MGP equation.  
Then, in the simpler case of a spherical trap we will compare 
different approaches in a situation where the number of atoms and the
frequency of the trap are kept fixed and the scattering length is allowed 
to vary in a representative range of values. This study is done to 
ascertain the degree of reliability of the MGP results.  
Finally, we will consider a cylindrical trap corresponding to the 
experimental situation. Depending on the value of the scattering length, 
the corrections to the GP results  can be as large as   
 30 $\%$ for energy, chemical potential and $\sim$40 $\%$ for 
the extracted scattering length.

In the spherical case, the energy functional associated with
 the MGP equation:
\begin{equation}
E_{MGP}[\psi] = \int d{\bf r} \left [ \frac {\hbar^2}{2 m} 
\mid \nabla \psi(r)\mid ^2 +
\frac {m}{2} \omega^2 r^2  \mid \psi \mid ^2+ \frac {2 \pi \hbar^2 a}{m} 
\mid \psi\mid ^4 + \frac {256 \hbar^2}{15 m} {\sqrt {a^5 \pi}}\mid \psi 
\mid ^5
\right ],
\label{eq:GPE}
\end{equation}
 is obtained in the local--density approximation by keeping the first 
two terms in the low density expansion for the energy 
density of a homogeneous system of hard-spheres, whose diameter
 coincides with the scattering length~\cite{Fetter71}:
\begin{equation}
\frac {E}{V}= \frac {2 \pi n^2 a \hbar^2 }{m} \left [ 
1+ \frac{128}{15} \left ( \frac
{na^3}{\pi} \right )^{1/2} + 8 (\frac {4}{3} \pi - \sqrt {3} ) (na^3) \ln (na^3)
 + O(na^3) \right ].
\label{eq:HHP}
\end {equation}
Up to this order of the expansion, the details of the potential do not
show up, and any potential with the same scattering length would give 
identical results. This universal behavior has recently been 
checked by a diffusion Monte Carlo calculation (DMC)~\cite{Boro99},  
providing the exact solution of the many--body Schr\"odinger equation.
 In Refs.~\cite{Fabro99,Boro99,Polls00} it was  shown that, 
for an uniform system, 
the first term of the expansion is accurate only at very low values of $x$, 
while the addition of the second term gives a good representation of the 
exact DMC results up to $x=10^{-2}$. The inclusion of the logarithmic 
term severely spoils the agreement already at intermediate $x$--values 
and therefore it has  not been incorporated into the functional energy, 
 $E_{MGP}[\psi]$. In the same references it was also shown that the energy 
functional computed in hypernetted chain (HNC) theory, explicitly
taking into account the interatomic correlations induced by the 
potential, provided a description very close to the MGP one.

It is convenient to simplify the notation by expressing lengths 
and energies in harmonic oscillator (HO) units. The
spatial coordinates, the energy, and the wave functions are rescaled as 
${\bf r}= a_{HO} {\bf r}_1$, $E= \hbar \omega E_1$, and 
$\Psi(r) = (N/a_{HO}^3)^{1/2} \Psi_1(r_1)$, where $\Psi_1(r_1)$ 
is normalized to unity and $a_{HO} = (\hbar/m \omega)^{1/2}$. 
 Using these new variables and performing a 
functional variation of $E_{MGP}[\psi]$, 
one gets the  modified Gross-Pitaevskii equation,
\begin{equation}
 \left [- \frac{1}{2} \nabla_{r_1}^2  +  
\frac{1}{2}  r_1^2 +
 4 \pi a_1 N \mid \psi_1(r_1) \mid^2 
 + {\sqrt {\pi a_1^5 N^3}} \frac {128}{3} 
\mid \psi_1(r_1)\mid ^3   
\right ] \psi_1(r_1)= \mu_1 \psi_1(r_1)~,
\end{equation}
where $a_1 = a/a_{HO}$ and $\mu_1$ is the chemical potential 
in HO units. The GP approximation is recovered by dropping the 
$\mid \psi_1(r_1)\mid ^3$ term.   

Table \ref{tab:tab1} gives some results for 
$N=10^4$ $^{85}$Rb atoms confined in a spherical trap with an oscillator 
angular frequency $\omega_{HO}/2 \pi = 
(\omega_\perp^2 \omega_z)^{1/3} /2 \pi= 12.77$ Hz, where 
$\omega_\perp /2 \pi = 17.5 $Hz and $\omega_z/2 \pi=6.9$ Hz are the 
radial and axial frequencies associated with the external potential of the 
cylindrical trap used in Ref.~\cite{Cornish1}. 
In the table we study the dependence
of the energy per particle and of the chemical potential on the 
scattering length, given in units of the Bohr radius of the Hydrogen atom, 
$a_0$. For this trap, $a_{HO}= 57657 a_0$. We also show the TF results. 
In this approximation it is often possible to derive simple analytical 
expressions~\cite{Edwards95}, useful to get quick estimates of several 
quantities. For instance, $\mu_1^{{\rm TF}}= 1/2 (15 a_1 N)^{2/5}$. 
Also reported are the peak values of the gas parameter, 
$x_{pk}= n(0)~a^3= N a_1^3 \mid \psi_1(0)\mid ^2$, whose TF 
estimate is $x_{pk}^{{\rm TF}}=( 15^2 a_1^{12} N^2 )^{1/5} /(8\pi)$.

At low values of the scattering length, MGP corrections are small and the
TF approach to the GP equation is not fully satisfactory. As expected,
the TF and GP results are much closer when $a$ increases and the
MGP corrections become important and of the order of $30\%$ at 
the largest value of $a= 10000 a_0$. $x_{pk}$ increases with $a$. 
The MGP density distribution gets wider and $x_{pk}^{{\rm MGP}}$ is 
depleted with respect to both TF and GP because of the 
repulsive character of the extra term in the MGP equation. The HNC 
results are confortably close to the MGP ones, supporting the use 
of only the latter approach in the remaininng of the paper. 

An analogous behavior is found in the anisotropic case. 
The results for the cylindric trap 
used in the experiments are given in Table \ref{tab:tab2}. 
Most of the TF results are again analytical.  
The HO units in the cylindric case are:
${\bf r}= a_{\perp,HO} {\bf r}_1$, $E= \hbar \omega_\perp E_1$, 
$\Psi({\bf r}) = (N/a_{\perp,HO}^3)^{1/2} \Psi_1({\bf r}_1)$, and 
$a_{\perp,HO} = (\hbar/m \omega_\perp,)^{1/2}$. The trap deformation 
parameter is $\lambda=\omega_z / \omega_\perp = 0.39$. 

An accessible experimental quantity connected to the density profile is the
already defined column density. Its TF expression is 
\begin{equation}
n_c^{TF}(z_1)= {\frac {2}{12 \pi a_1 N}} \left [2 (\mu_1^{TF} - 
{\frac {1}{2}} \lambda^ 2 z_1^2 ) \right ]^{3/2}~.
\end{equation}
A measure of the extension of the condensate is the half maximum radius of
the column density, $R_{1/2}$, defined as the $z_1$ value where 
$n_c(z_1=R_{1/2})=\frac {1}{2} n_c(0)$. Also interesting is the full 
strength at half maximum, FSHM, given by the 
integrated strength of the column between the $\pm R_{1/2}$ values.

\begin{figure}
\epsfig{file=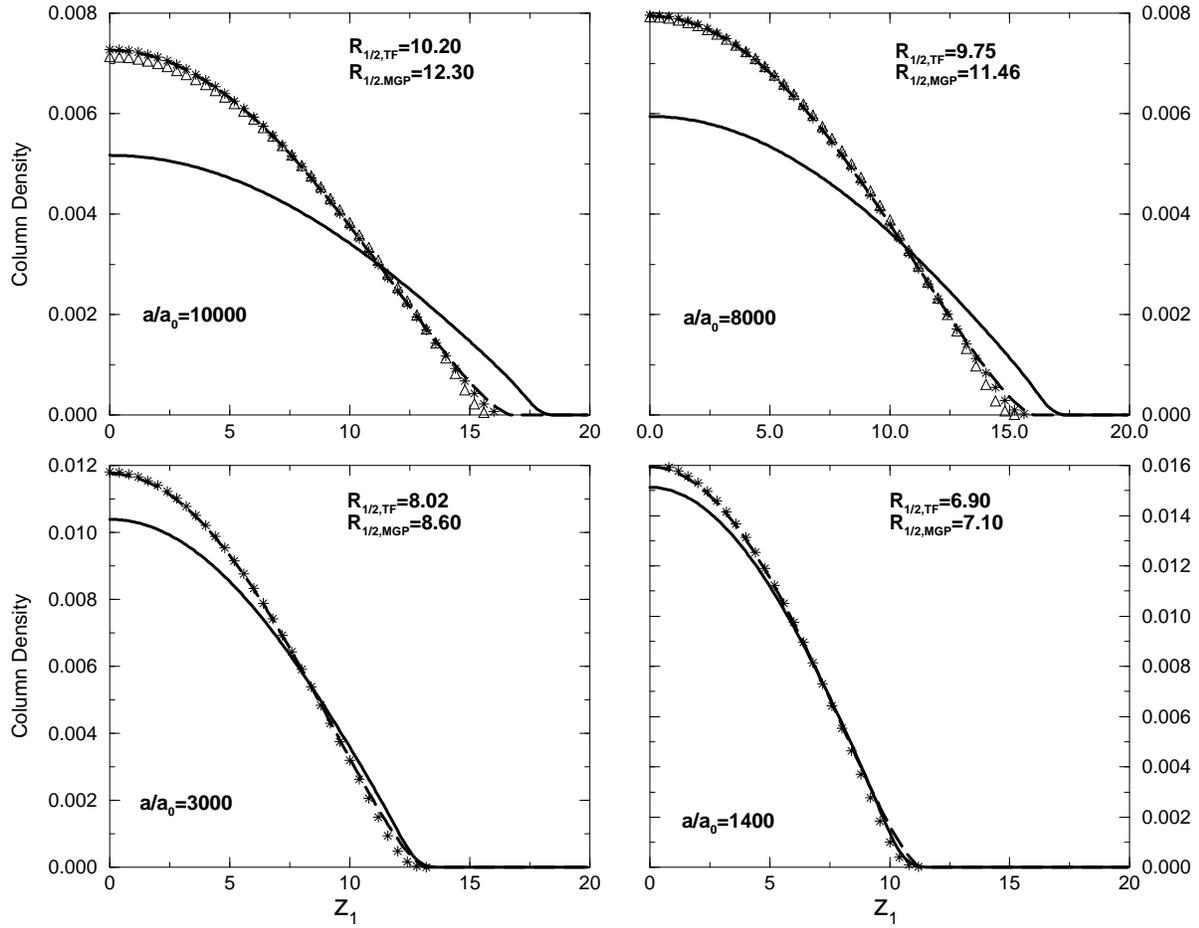,width=15 cm}
\caption{ 
Column densities at four values of the scattering length for the cylindrical 
trap. Dashed lines= TF, stars= GP, solid lines= MGP. The triangles in the 
first (second) upper panel give the MGP column density at $a/a_0$=5920 (4940).
} 
\label{fig:fig1}
\end{figure}

In Fig. \ref{fig:fig1}, we show the column densities in different approaches, 
for the same  set of scattering lengths reported in the 
Tables.  The solid and dashed lines correspond to the MGP and TF 
results, respectively; stars are the GP densities; the triangles in the 
two upper panels give $n_c(z_1)$ evaluated in MGP, but changing the  
scattering length to reproduce $R_{1/2,TF}$, supposedly corresponding to the 
measured radius.  The two values are $a/a_0=5920$ for R$_{1/2,TF}$=10.20 
and $a/a_0=4940$ for $R_{1/2,TF}$=9.75 and the related  MGP columns 
are practically identical to the TF ones. 

The GP and TF results almost coincide and the MGP corrections 
are more sizeable at the two largest values of $a$, where $x_{pk}$ becomes 
of the order of 10$^{-2}$. Because of the repulsive nature of the MGP extra 
term, $R_{1/2,MGP}$ is larger than $R_{1/2,TF}$, while 
FSHM$_{MGP}$ is smaller than FSHM$_{TF}$, and for low $z_1$--values 
$n_c^{MGP}(z_1)$ lies  below $n_c^{TF}(z_1)$. A smaller scattering length is 
required to reproduce $R_{1/2,TF}$ and FSHM$_{TF}$ in the MGP approach 
in the high $x_{pk}$ region. 
In fact, we find a reduction of $\sim 40 \%$ of $a/a_0$. 
This analysis shows that using the TF column density to extract the 
scattering length in the large gas parameter regime could lead to severe  
overestimates in this kind of trap geometry.

The lower panels of the figure roughly correspond to 
$x_{pk}\sim$10$^{-3}$--10$^{-4}$. As expected, the MGP corrections 
are smaller and the computed $R_{1/2}$ values become closer 
when $a/a_0$ decreases.    

\begin{figure}
\epsfig{file=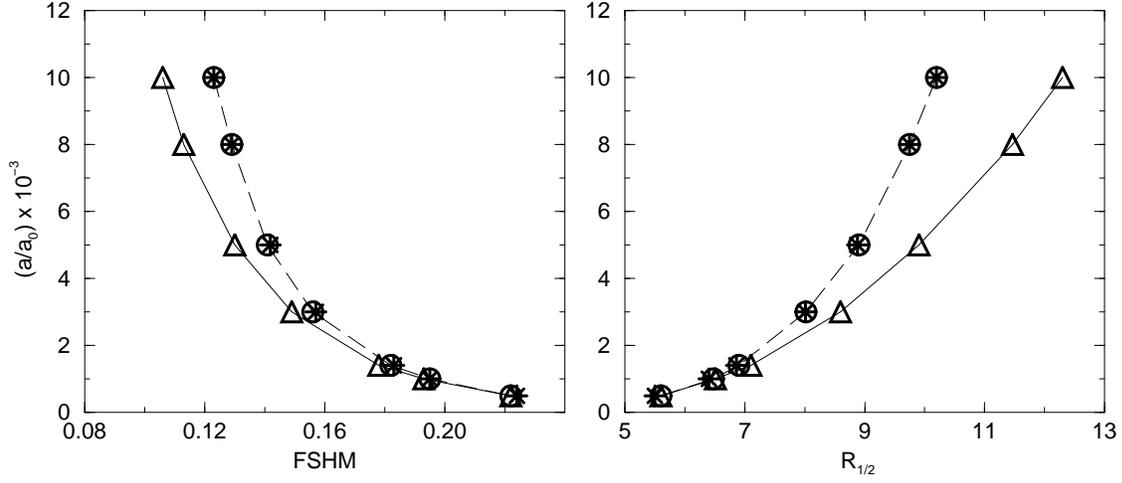,width=15 cm}
\caption{ 
Scattering length as a function of the full strength at half maximum 
(left) and of the half maximum radius (right) in the cylindrical trap. 
Circles, stars and triangles correspond to the TF, GP and MGP results, 
respectively. Lines are a guide to the eyes. 
} 
\label{fig:fig2}
\end{figure}

Fig. \ref{fig:fig2} shows the scattering length as a function of 
FSHM and $R_{1/2}$  for the cylindrical trap within the three 
methods we have analyzed. 
The figure stresses that, depending on the FSHM and $R_{1/2}$ values and 
on the approach, the estimates of $a$ can differ by up to $40 \%$.

In conclusion, we find that the MGP equation induces corrections of 
$30 \%$ in the ground state properties of the condensate, 
when the conditions of the JILA experiments for $^{85}$Rb are considered. 
Comparable corrections are obtained for the column densities, where 
large differences between the MGP and the standard TF and GP results may 
be found. 
These differences appear to be relevant for the extraction of the scattering 
length when large values of the gas parameter come into play. MGP is  still 
a mean field theory, since it tries to incorporate correlation effects 
into the average single particle potential. However, we believe that its 
predictions are probably indicative at those regimes attained in recent 
experiments. 

Useful discussions with Ennio Arimondo are gratefully acknowledged. 
This research was partially supported by DGICYT (Spain) Grant
No. PB98-1247, the agreement CICYT (Spain)--INFN (Italy),  
the program SGR98-11 from Generalitat de Catalunya and the 
Progetto di Ricerca di Interesse Nazionale: {\em Fisica Teorica del
Nucleo Atomico e dei Sistemi a Molticorpi} from MURST.

\newpage 
 
\begin{table}               
\caption{ 
Ground state properties of $N=10^4$ $^{85}$Rb 
atoms confined in a spherical trap ($\omega/2\pi$=12.77 Hz) 
in different approaches. 
$\mu_1$=chemical potential, $E_1/N$=energy per atom, $x_{pk}$=peak gas 
parameter. Energies in HO units. 
}
\begin{tabular}{c|cccc}
 $a/a_0$ & 1400 &  3000 & 8000 & 10000 \\
\tableline
 $\mu_1^{TF}$  & 13.29   & 18.02  & 26.00 & 29.18 \\
 $\mu_1^{GP}$  & 13.41   & 18.12  & 26.75 & 29.24 \\
 $\mu_1^{MGP}$ & 13.95   & 19.82  & 33.34 & 38.01 \\
 $\mu_1^{HNC}$ & 13.90   & 19.66  & 33.33 & 38.41  \\
\tableline
 $E_1^{TF}/N$  & 9.50    & 12.87  & 18.57 & 20.84 \\
 $E_1^{GP}/N$  & 9.66    & 13.01  & 19.16 & 20.93 \\
 $E_1^{MGP}/N$ & 10.00   & 14.09  & 23.40 & 26.60 \\
 $E_1^{HNC}/N$ & 9.97    & 13.98  & 23.24 & 26.59  \\
\tableline
 $x_{pk}^{TF}$ & 6.23 $\times 10^{-4}$ & 3.88 $\times 10^{-3}$    & 
 4.09 $\times 10^{-2}$ & 6.98 $\times 10^{-2}$ \\ 
 $x_{pk}^{GP}$ &   6.26 $\times 10^{-4}$ & 3.89 $\times 10^{-3}$ & 4.09
 $\times 10^{-2}$ & 6.99 $\times 10^{-2}$ \\
 $x_{pk}^{MGP}$ &   5.70 $\times 10^{-4}$ & 3.18 $\times 10^{-3}$ & 2.59 
 $\times 10^{-2}$ & 4.10 $\times 10^{-2}$ \\
 $x_{pk}^{HNC}$& 5.75 $\times 10^{-4}$  & 3.24 $\times 10^{-3}$  & 2.52 
 $\times 10^{-2}$ & 3.85 $\times 10^{-2}$       \\
\end{tabular}
\label{tab:tab1}
\end{table}

\begin{table}               
\caption{ 
Ground state properties of $N=10^4$ $^{85}$Rb 
atoms in the cylindrical trap described in the paper. Energies in HO units. 
}
\begin{tabular}{c|cccc}
 $a/a_0$ & 1400 &  3000 & 8000 & 10000 \\
\tableline
 $\mu_1^{TF}$  &  9.70 & 13.15 & 19.47 & 21.29 \\
 $\mu_1^{GP}$  &  9.82 & 13.25 & 19.55 & 21.36 \\
 $\mu_1^{MGP}$ & 10.22 & 14.51 & 24.38 & 27.79 \\
\tableline
 $E_1^{TF}/N$  &  6.93 & 9.39  & 13.91 & 15.21 \\
 $E_1^{GP}/N$  &  7.08 & 9.52  & 14.00 & 15.29 \\
 $E_1^{MGP}/N$ &  7.33 & 10.31 & 17.09 & 19.43 \\
\tableline
 $x_{pk}^{TF}$ & 6.23 $\times 10^{-4}$ & 3.88 $\times 10^{-3}$ &
  4.09 $\times 10^{-2}$ & 6.98 $\times 10^{-2}$ \\
 $x_{pk}^{GP}$ & 6.28 $\times 10^{-4}$ & 3.90 $\times 10^{-3}$ &
  4.10 $\times 10^{-2}$ & 7.00 $\times 10^{-2}$ \\
 $x_{pk}^{MGP}$& 5.72 $\times 10^{-4}$ & 3.19 $\times 10^{-3}$ & 
  2.60 $\times 10^{-2}$ & 4.10 $\times 10^{-2}$ \\
\end{tabular}
\label{tab:tab2}
\end{table}

\end{document}